# Fabrication of an integrated high-quality-factor (high-Q) optofluidic sensor by femtosecond laser micromachining

Jiangxin Song[a,b], Jintian Lin[a,b], Jialei Tang[a,b], Yang Liao[a], Fei He[a], Zhaohui Wang[a,b], Lingling Qiao[a], Koji Sugioka[c], Ya Cheng[a,*]

[a] *State Key Laboratory of High Field Laser Physics, Shanghai Institute of Optics and Fine Mechanics, Chinese Academy of Sciences, Shanghai 201800, P. R. China.*
*Fax: +86 6991 8021; Tel: +86 6991 8546; E-mail: ya.cheng@siom.ac.cn*

[b] *Graduate School of Chinese Academics of Science, Beijing 100039, China*

[c] *Center for Advanced Photonics, RIKEN, Hirosawa 2-1, Wako, Saitama 351-0198, Japan.*

We report on fabrication of a microtoroid resonator of a high-quality factor (i. e., Q-factor of ~$3.24\times10^6$ measured under the critical coupling condition) integrated in a microfluidic channel using femtosecond laser three-dimensional (3D) micromachining. Coupling of light into and out of the microresonator has been realized with a fiber taper that is reliably assembled with the microtoroid. The assembly of the fiber to the microtoroid is achieved by welding the fiber taper onto the sidewall of the microtoroid using $CO_2$ laser irradiation. The integrated microresonator maintains a high Q-factor of $3.21\times10^5$ as measured in air, which should still be sufficient for many sensing applications. We test the functionality of the integrated optofluidic sensor by performing bulk refractive index sensing of purified water doped with tiny amount of salt. It is shown that a detection limit of ~$1.2\times10^{-4}$ refractive index unit can be achieved. Our result showcases the capability of integration of high-Q microresonators with complex microfluidic systems using femtosecond laser 3D micromachining.

## Introduction

Recently, femtosecond laser direct writing has emerged as a straightforward approach for fabrication of three-dimensional (3D) microfluidic structures and integrated optofluidic devices in glass with great flexibility [1-2]. By focusing a femtosecond laser beam in glass, the interior of glass can be modified in a spatially selective manner through multiphoton absorption. The laser-irradiated regions in glass can be selectively etched by subsequent wet etching using aqueous solutions of etchants such as hydrofluoric (HF) acid, allowing for formation of microfluidic systems of complex 3D configurations directly in glass [3-5]. Likewise, free-space optical components of high optical properties such as micromirrors and microlenses can also be produced in glass [6-8]. In addition, irradtion with tightly focused femtosecond laser also induces refractive index increase in glass. Thus, the same femtosecond laser used for fabricating microfluidic structures can also be employed for optical waveguides writing [9-10]. By taking all of these advantages, many novel optofluidic devices featured by their 3D layouts have been built, including optofluidic senors, cell sorters, nanoaquaria, etc [11-14].

For many optofluidic sensing applications, efficient enhancement of the sensing sensitivity can be achieved by use of a high-quality factor (Q factor) optical cavity [15-17]. In particular, it has been shown recently that a whispering gallary mode (WGM) microresonator can be used for biosensing with a detection limit down to single molecules, owing to its unique capability of confining light in a tiny volume for long periods of time by total internal reflection [18-19]. The WGM microresonators demonstrated in these experiments are typically microspheres or microtoroids. In most cases, coupling of light into and out of the resonators is realized with a fiber taper carefully positioned near the microresonator with precision motion stages to achieve the critical coupling condition [20,21]. Since the fiber taper and the microresonators are not integrated, the widespread use of such systems in optofluidic and biosensing applications has been hampered because of their poor portbility. Further chip-level integration of the microresonators in a microfluidic system is also difficult, because of the incompatibility between the processing techniques of microfluidic systems and optical microresonators.

As we have demonstrated in 2012, WGM microresonators of Q-factors on the order of $10^6$ and of 3D geometries can be fabricated in fused silica using femtosecond laser direct writing followed by chemical wet etching, i.e., the same approach that has been employed for fabrication of microfluidics in fused silica [22]. The unique advantage facilitates creation of fully integrated optofluidic sensors with a high-Q resonator directly incorporated in microfluidic systems, as both kinds of structures can be fabricated simultaneously in glass chips. In this work, we demonstrate such an optofluidic sensor with detection limit of $1.2\times10^{-4}$ refractive index unit (RIU) for bulk refractive index sensing of liquids. It should be specifically noted that in the fabricated device, a fiber taper is reliably assembled with the microresonator by $CO_2$ laser welding for coupling light into and out of the resonator. The fully chip-level integrated device can be useful for label-free field portable sensing applications.

## Process for fabrication of the integrated optofluidic sensor

### Fabrication of microfluidic channels and microresonator

In this work, commercially available fused silica glass substrates (UV grade fused silica JGS1 whose upper and bottom surfaces were polished to optical grade) with a thickness of 1 mm were used. In our previous works, the processes of fabrication of both microfluidic channels and microtoroid resonators have been established, in which the details can be found [1,22,23]. Here, we only briefly describe the fabrication process.

The process flow for fabrication of the fully integrated sensor mainly consists of three steps: (1) femtosecond laser exposure followed by selective wet etch of the irradiated areas to create the 3D microfluidic channels embedded in glass and the microdisk structure at the outlet of the microchannel; (2) selective reflow of the silica disk by $CO_2$ laser irradiation to improve the Q-factors; and (3) assembly of the fiber taper with the microresonator by $CO_2$ laser welding.

The femtosecond laser system consists of a Ti: sapphire oscillator (Coherent, Inc.) and a regenerative amplifier, which emits 800 nm, ~40 fs pulses with a maximum pulse energy of ~4.5 mJ at 1-kHz repetition rate. The initial 8.8-mm-diameter beam was reduced to 5 mm by passing through a circular aperture to guarantee a high beam quality. Power adjustment was realized using neutral density (ND) filters. The glass samples could be arbitrarily translated in 3D space with a resolution of 1 μm by a PC-controlled XYZ stage combined with a nano-positioning stage.

To fabricate the microfluidic channels, the femtosecond laser beam was focused 300 μm below the glass surface with a 20× objective lens (NA=0.46) and scanned at a speed of 50 μm/s. After the direct writing of the microchannels in the glass, the sample was soaked in a solution of 10% HF acid diluted with water to selectively remove the materials in the region modified by femtosecond laser irradiation. The chemical etching was carried out in an ultrasonic bath for 3 hrs. It should be mentioned here that an open reservoir connecting the microfluidic mixer, which was formed in the intersecting area of two open microfluidic channels as shown in Figs. 1(a) and (b), would be fabricated later together with the microdisk. Thus, the region of the reservoir was not irradiated with the femtosecond laser pulses at this stage, and consequently, no selective etching could occur in this region, too.

The microtoroid resonator was formed together with the open reservoir after the 3D microfluidic channels had been constructed in the glass due to the much shorter time required in the chemical etching to form the open channels and the microdisk structure. To fabricate the reservoir, the same irradiation conditions as that used for fabrication of the 3D microfluidic channels was chosen. However, during the irradiation, we intentionally avoided a square-shaped area (150 μm×150 μm) in the center of the microreservoir which was reserved for the subsequent fabrication of the microdisk. The microdisk supported by a thin pillar was fabricated by a layer-by-layer annular scanning with a nano-positioning stage of the lateral scanning step set to be 1 μm, and the scanning speed was set at 600 μm/s. A 100× objective with a numerical aperture (NA) of 0.8 was used to focus the beam to a ~1 μm-dia. spot, and the average femtosecond laser power measured before the objective was ~0.05 mW. After the laser writing, the sample was etched with an HF acid diluted in water at a concentration of 5% for 0.5 hrs. At this stage, the sidewall of the microdisk had a roughness of several tens of nanometer, which is not suitable for high-Q resonator applications.

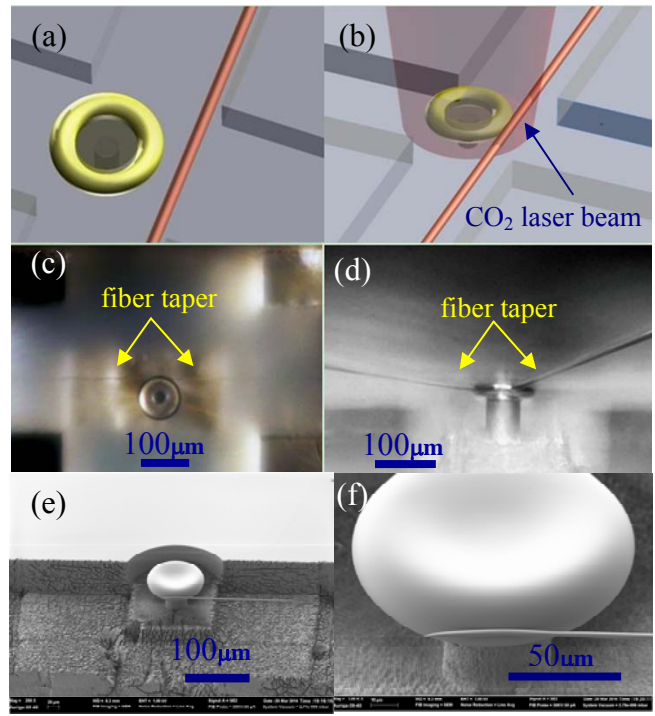

Fig.1 (a-b) Schematic illustration of assembling a fiber taper onto the microresonator. (a) Bring the fiber taper close to the microresonator to achieve a high coupling efficiency. (b) Bring the fiber taper in contact with the microresonator and perform $CO_2$ laser welding. (c) Top-view optical micrograph of the fiber taper coupled to the microresonator before the welding. (d) Side-view optical micrograph of the fiber taper welded onto the sidewall. Fiber can remain assembled on the microtoroid with bending. Note that the microtoroid is fabricated in the intersecting area of two open channels using the femtosecond laser. Overview (e) and close up view (f) SEM images of the fiber taper welded to the sidewall of microtoroid.

To improve the surface smoothness on the sidewall of the microdisk and achieve the desirable high Q-factor, we smoothed the surface of the microdisk by introducing a surface-normal-irradiation with a $CO_2$ laser (Synrad Firestar V30). The $CO_2$ laser beam was focused by a lens to a circular spot of approximately ~100 μm in diameter. The laser was operated with a repetition rate of 5 kHz, and the irradiation strength was controlled by adjusting the duty ratio. As the disk diameter thermally shrinked during reflow, surface tension induced a collapse of the fused silica disk, leading to a toroidal-shaped boundary. During this process the disk was monitored by a charge coupled device (CCD) from the side using a 20× objective lens. The total reflow process took merely 4 s at a duty ratio of 5.0%. Due to the surface tension, the smoothness of the microtoroid surface is excellent, as shown in the optical micrograph [Figs. 1(c) and (d)]. The microtoroid diameter was measured to be ~80μm with a 10-μm-thick toroid-shaped boundary.

To characterize the mode structure and Q factor of the microtoroid cavity, resonance spectra were measured via the optical fiber taper coupling method. A swept-wavelength tunable external-cavity New Focus diode Laser (Model: 6528-LN) and a dBm Optics swept spectrometer (Model:4650) were used to measure the transmission spectrum from the fiber taper with a resolution of 0.1 pm. The fiber taper was fabricated by heating and stretching a section of optical fiber (SMF-28, Corning) until reaching a minimum waist diameter of approximately ~1 μm. The fiber taper can provide an evanescent excitation of whispering gallery modes of the cavity. The fabricated optofluidic sensor was fixed on a three-axis nano-positioning stage with a spatial resolution of 50-nm in the XYZ directions. We used dual CCD cameras to simultaneously image microtoroid cavity and fiber taper from the side and the top [22].

**Assembling of fiber taper onto the microresonator**

The key innovation of this experiment is to assemble the fiber taper onto the sidewall of the microtoroid resonator to produce a fully integrated, portable optofluidic sensor [24]. Figure 1 illustrates the procedures of assembling of the optical fiber taper with the resonator. As illustrated in Fig. 1(a), the fiber taper was first brought to the vicinity of the toroid using the 3D nano-positioning stage to achieve the critical coupling. The highest Q-factor we achieved under the critical coupling condition was $3.24\times10^6$, as evidenced by the transmission spectrum provided in Fig. 2. Then we slightly moved the fiber taper to have it in direct contact with the sidewall of the microtoroid, and irradiated the microtoroid with the $CO_2$ laser beam again, as illustrated in Fig. 1(b). We notice here that the focusing conditions of the $CO_2$ laser beam used for welding the fiber onto the microtoroid were the same as that used for reflow the microdisk to form the microtoroid, namely, no modification on the optical setup which was established for surface reflow of the microresonator was applied. However, to achieve a high Q-factor of the assembled microtoroid-fibre system, we indeed had optimized the duty ratio of the $CO_2$ laser. In our experiment, eventually we chose a duty ratio of 2.5% (i.e., 50% lower than that used for reflowing the microdisk), leading to both a strong welding and a decent Q-factor in air of $3.21\times10^5$ (see Fig. 3 below).

Figure 1(c) show the optical micrograph of the microtoroid in contact with the fiber taper, before and after the fiber taper being welded onto its sidewall with the $CO_2$ laser irradiation, respectively. Specifically, in Fig. 1(d), one can see that the $CO_2$ laser irradition creates a strong welding in the contacting area between the fiber taper and the microtoroid, thus the fiber can be significantly bent without falling apart from the microtoroid. The high mechanical strength will ensure a reliable operation of the sensor in field-portable uses. Figures 1(e) and (f) further provide detailed information on how the fiber taper is welded to the sidewall with the respective overview and close-up view scanning electron microscope (SEM) images of the integrated fiber-microtoroid system. It can be seen that after the welding, the cross section of the fiber taper is significantly deformed, i.e., the original perfectly round-shaped cross section becomes compressed flat oblate. Further theoretical analysis is required to understand how the deformation will influence the Q factor in a quantitative way.

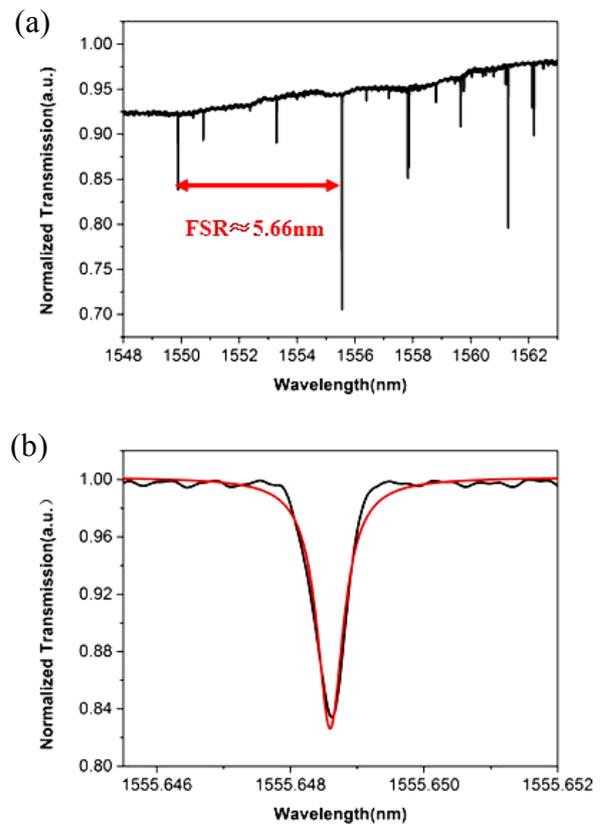

Fig. 2 (a) Transmission spectrum of the microtoroid measured under the critical coupling condition. (b) The Rorentz fitting (red curve) of one dip showing a Q-factor of $3.24\times10^6$.

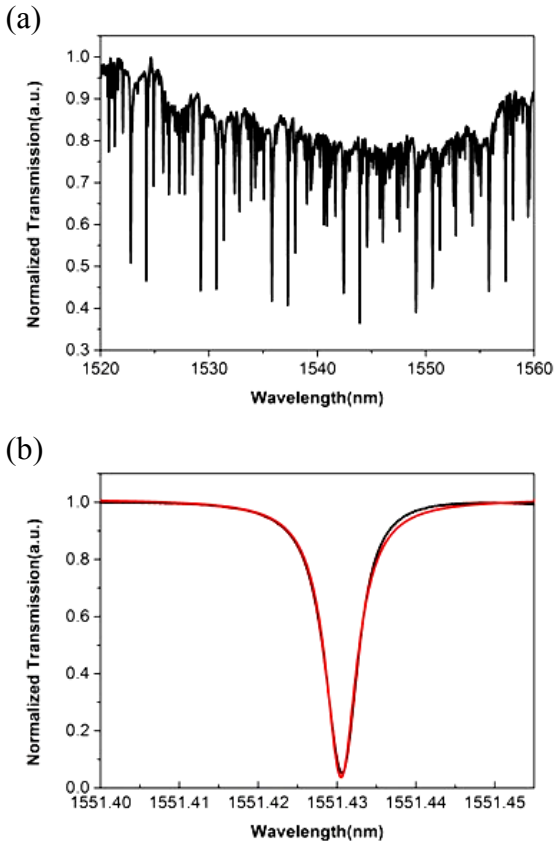

Fig. 3 (a) Transmission spectrum of the integrated microtoroid-fiber system measured in air after the $CO_2$ laser welding. (b) The Rorentz fitting (red curve) of one dip showing a Q-factor of $3.21 \times 10^5$.

## Results and discussion

**Characterization of integrated microtoroid-fiber system**

An important question is that whether the integrated toroid-fiber system can maintain a decent Q-factor for high-sensitivity sensing applications after the $CO_2$ laser welding. Figure 2(a) shows the transmission spectrum of the femtosecond laser fabricated microtoroid measured under the critical coupling condition. The free spectral range of 5.66 nm agrees well with the theoretical calculation. As shown in Fig. 2(b), the Q-factor calculated based on a Lorentz fitting reaches $3.24 \times 10^6$, which is consistent with our previous results [22,23].

For comparison, the transmission spectrum of the integrated microtoroid-fiber system is shown in Fig. 3(a). The spectrum shows more complicated spectral structure mainly because of the excitation of high-order modes. Again, based on the Lorentz fitting as shown by the red curve in Fig. 3(b), the Q-factor is estimated to be $\sim 3.21 \times 10^5$. This is significantly lower than that achieved with the critical coupling, however, for many sensing applications, such a value of Q-factor can already ensure high detection limits.

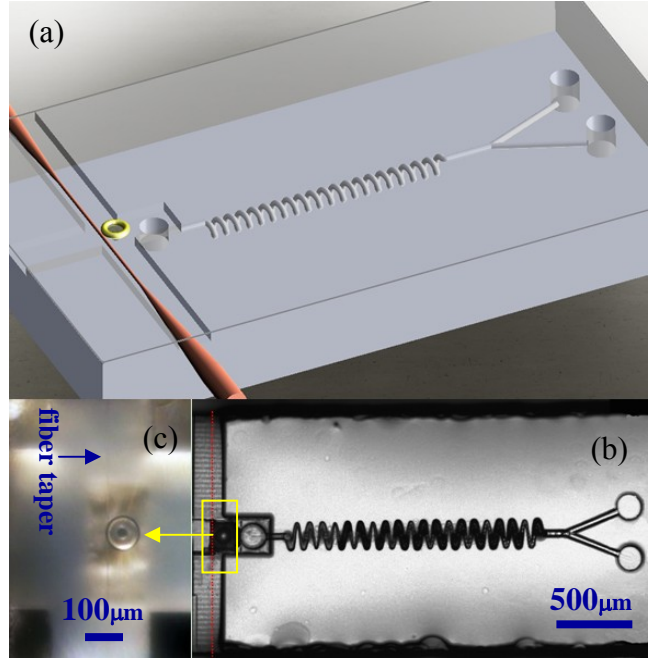

Fig. 4 (a) Illustration of the layout of the integrated optofluidic sensor. (b) The optical micrograph of the fabricated sensor. The fiber is indicated with the dashed red line, and the microtoroid resonator assembled with the fiber taper is sitting in the yellow square. (c) The close-up view of the assembled microtoroid-fiber system fabricated in the microfluidic channel using a femtosecond laser.

**Refractive index sensing based on the integrated sensor**

At last, we demonstrate a fully integrated optofluidic sensor in which a microfluidic system and the microtoroid resonator were fabricated using femtosecond laser direct writing and the assembing of fiber taper was achieved based on the procedures described in Figs. 1(a) and 1(b). Figure 4(a) illustrates the layout of the designed sensor. The microfluidic system is embedded 300 μm beneath the glass surface, which is composed of two Y-branched-channels with a length of 600 μm connected with two inlets, and a long spiral channel with a total length of 3 mm and a diameter of ~35 μm connected to an outlet for mixing of liquid. Near the outlet, a reservoir was fabricated and the microtoroid resonator was fabricated in the center area of reservoir. Therefore, the mixed liquid coming out from the outlet can be sensed with the microresonator. The optical microscope image of the fabricated sensor is presented in Fig. 4(b), showing that all the components have been integrated in a single glass chip. The close-up view image of the integrated toroid-fiber system is shown in Fig. 4(c). The fibre taper is too thin and can hardly be seen in such images of limited resolutions (Figs. 4(b) and (c)) which are for providing a sufficiently large field-of-view for the sensor. Therefore, the fiber taper is indicated with the red dashed line in Fig. 4(b), and with an arrow in Fig. 4(c).

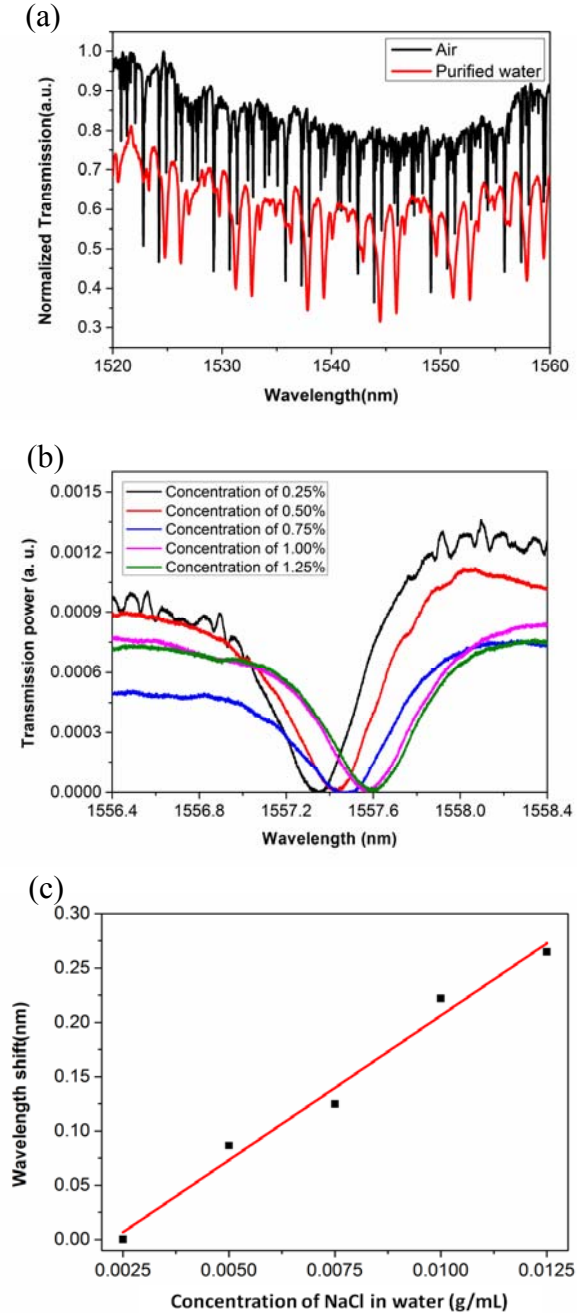

Fig. 5 (a) Transmission spectrum of the integrated microtoroid-fiber system measured in air after the $CO_2$ laser welding (black curve) and that of the integrated microtoroid-fiber system measured in purified water (red curve). (b) The close-up view of the spectral dips around 1557 nm wavelength measured at different doping concentrations of salt. (c) The wavelength shift plotted as a function of the doping concentration of salt (black squares), which can be well fitted with a linear relationship (red curve).

Figure 5 shows the results of sensing the refractive index of purified water doped with salt at different concentrations. Specifically, the overall transmission spectra of the microresonator in air and in purified water are compared by the black and red curves in Fig. 5(a), respectively. It can be seen that as compared to the spectra measured in air, the spectral width of each dip becomes larger, and the number of modes in the spectrum measured in water has been significantly reduced, mainly because of the reduced Q-factor and the ellimination of the high-order modes, respectively. The measured Q-factor of the microresonator immersed in purified water was only $\sim 3.1\times 10^3$. The severe reduction of the Q-factor in water is mainly caused by two factors. The first limiting factor is that the experiment was carried out at a wavelength of ~1550 nm (i.e., the tunable laser source in our lab), which is highly absorptive in water [25]. The second is that because of the limited range of motion of our nano-positioning stage, the size of our fabricated microtoroid resonator is about 80 μm, which is not large enough to support sufficiently high Q-factors in water as have been investigated before. Actually, in Ref. [26], the authors have clearly shown that for a fused silica microtoroid with a diameter of ~80 μm, it exihibits a Q-factor on the level of $\sim 10^3$ in water at in the spectral range around 1550 nm. The measured Q-factor of our fabricated resonator ($Q \sim 3.1 \times 10^3$) agrees well with their analysis.

Figure 5(b) shows five spectra recorded in purified water doped with salt at different concentrations of 0.25%, 0.5%, 0.75%, 1%, and 1.25%. Here, all the concentrations are given in weight percent wt.%. One can see that the position of the spectral dip around 1557 nm wavelength shifts significantly with the increasing concentration of the salt, which can be attributed to the change in the refractive index of the water with varying doping concentration. In addition, figure 5(c) shows that the position of the spectral dip shifts almost linearly with the increasing doping concentration of salt, which is reasonable because at such a low concentration of the salt, the refractive index of the liquid will also vary linearly with the doping concentration. Here, we note that as a premilary experiment, we did not use the microfluidic mixer fabricated in glass for preparing the liquid sample, but directly added a drop of the prepared liquid sample onto the microresonator. Nevertheless, there is no significant technical difficulty in using the fully integrated optofluidic sensor for those labs equipped with microfluidic instruments such as microfluidic pumps and so on.

Based on the data in Fig. 5(c), we have evaluated that the sensitivity of our sensor can reach ~220 nm/RIU, yielding a dection limit of $1.2\times 10^{-4}$ RIU based on the analysis method provided in Ref. [27]. The results are comparable to the best sensors fabricated by the femtosecond laser direct writing so far. However, we expect that there will be a large room for improvement of the detection limit of our integrated sensor. Assuming a Q-factor of $\sim 3\times 10^5$ of the microresonator in liquids which could be achieved with a larger-sized microtoroid and a detection light in the visible range, the dection limit could in principle reach $10^{-6} \sim 10^{-7}$ RIU for refractive index sensing [26]. It should be pointed out that in such high-sensitivity measurements, the temperature in the sensor must be well controlled for avoiding the refractive index fluctuation induced by temperature variations [11].

## Conclusions

In conclusion, we have demonstrated fabrication of a fully integrated optofluidic sensor with a microtoroid WGM resonator directly incorporated in a microfluidic system and a fiber taper reliably assembled onto the microresonator. Benefited from the 3D fabrication capability of femtosecond laser direct writing, the microfluidic channels embedded in glass and the microtoroid resonator supported by the glass stem can be automatically integrated in the laser writing process without post-assembly and bonding procedures. The microtoroid resonator exhibits a Q-factor of $3.24 \times 10^6$ in air under the critical coupling condition. After the fiber taper is welded onto the sidewall of the resonator by $CO_2$ laser welding, the Q-factor of the integrated system measured in air decreases to $\sim 3.21 \times 10^5$. Limited by the range of motion of our nano-positioning stage for fabrication of the microtoroid resonator, the size of the fabricated resonator is only 80 μm, which causes the limited Q-factor when the resonator is immersed in water and tested with a light at the wavelengths near 1550 nm. The detection limit presented in our work is at the same level of the best results demonstrated by other groups with the sensors fabricated by the femtosecond laser direct writing. However, by fabricating microtoroid resonators of larger sizes and reducing the absorption in liquids with detction light in the visible range, the microtoroid-based sensor can potentially offer much greater detection limit garanteed by the high-Q nature of the GWM resonators.

## Acknowlegements

This research is financially supported by National Basic Research Program of China (Nos. 2011CB808100 and 2014CB921300), National Natural Science Foundation of China (Nos. 60921004, 61275205, 61108015, and 11104294), and the Program of Shanghai Subject Chief Scientist (11XD1405500).

.